\documentclass[showpacs,preprint]{revtex4}
\usepackage{graphicx}
\usepackage{subfigure}
\usepackage{amssymb}
\usepackage{amsmath}
\usepackage{bm}
\usepackage{latexsym}
\usepackage{hyperref}

\begin {document}

\title{Dynamics of two qubits in a spin-bath with anisotropic XY coupling}
\author{Jun Jing and Zhi-Guo L\"{u}\footnote{Email address: zglv@sjtu.edu.cn}}
\affiliation{Department of Physics, Shanghai Jiao Tong University,
Shanghai 200240, China}
\date{\today}

\begin{abstract}
The dynamics of two 1/2-spin qubits under the influence of a quantum
Heisenberg $XY$ type spin-bath is studied. After the
Holstein-Primakoff transformation, a novel numerical polynomial
scheme is used to give the time-evolution calculation of the center
qubits initially prepared in a product state or a Bell state. Then
the concurrence of the two qubits, the $z$-component moment of
either of the subsystem spins and the fidelity of the subsystem are
shown, which exhibit sensitive dependence on the anisotropic
parameter, the temperature, the coupling strength and the initial
state. It is found that (i) the larger the anisotropic parameter
$\gamma$, the bigger the probability of maintaining the initial
state of the two qubits; (ii) with increasing temperature $T$, the
bath plays a more strong destroy effect on the dynamics of the
subsystem, so does the interaction $g_0$ between the subsystem and
the bath; (iii) the time evolution of the subsystem is dependent on
the initial state. The revival of the concurrence does not always
means the restore of the state. Further, the dynamical properties of
the subsystem should be judged by the combination of concurrence and
fidelity.

\end{abstract}

\maketitle

\section{Introduction}

Solid-state devices, in particular, ultra-small quantum dots
\cite{Burkard} with spin degrees of freedom embedded in
nanostructured materials, compared with other physical systems such
as ions in trap \cite{Sackett}, are more easily scaled up to large
registers and they can be manipulated by energy bias and tunneling
potentials \cite{Loss}. The key building block of a quantum
processor consists of two entangled quantum bits. Thus the spin
system is one of most promising candidates for quantum computation
owing to long relaxation and decoherence times \cite{Loss, Kane,
Nielsen}. However, the spin qubits are open systems which is
impossible to avoid interactions with their environments
\cite{Breuer2, weiss, Breuer, Yuan}. Finally, the states of the
qubits will relax into a set of ``pointer states'' in the Hilbert
space \cite{Zurek1981}; and the entanglement between the spin qubits
will also vanish. Yet the entanglement is the most intriguing
feature of quantum composite system and the vital resource for
quantum computation and quantum communication \cite{Nielsen,
Bennett}. These are so-called decoherence and disentanglement
processes. These two disadvantages will not be overcome until the
modelling of the surrounding environment or
bath of the spin systems. \\

For solid state spin nano-devices, the quantum noise mainly arises
from the contribution of nuclear spins, which could be regarded as a
spin environment. Recently, there are lots of works were devoted to
study the behavior of center spins under the strong non-Markovian
influence of a spin-bath \cite{Stamp, Loss2}. Lucamarini and
co-workers made use of perturbative expansion method
\cite{Paganelli} and mean-field approximation \cite{Paganelli2} to
study the temporal evolution of entanglement pertaining to qubits
interacting with a thermal bath. They found entangled states with an
exponential decay of the quantum correlation at finite temperature.
Hutton and Bose \cite{Hutton} investigated a star network of spins
at zero temperature, in which all spins interact exclusively and
continuously with a central spin through Heisenberg XX couplings of
equal strength. Their work was advanced by Hamdouni and co-workers
\cite{Hamdouni}, who derived the exact reduced dynamics of a central
two-qubit subsystem in the same bath configuration. And they also
studied the entanglement evolution of the central system. Yuan and
co-workers \cite{Yuan} used a novel operator technique to obtain the
dynamics of the two coupled spins in quantum Heisenberg XY high
symmetry spin model. The results of all the above works are very
exciting. Yet their methods are of some kinds of complex analytical
derivations. And in Ref. \cite{Yuan}, their analytical results are
dependent on some particular initial states and essentially the
interaction between the spins in their model is isotropic. Here we
introduce a ``half analytical and half numerical'' method to solve
such an open quantum system problem in an anisotropic Heisenberg XY
model. The present model involves the Heisenberg XY interaction that
has broad applications for various quantum information processing
systems, such as quantum dots, Cavity-QED, etc\cite{Imamoglu, Zheng,
Wang, Lidar}. Besides, our method is initial states independent.\\

In this paper, we study an open two-spin-qubit system in a spin
bath of star-like configuration, which is similar to the cases
studied in Ref. \cite{Yuan, Hamdouni}. But the two qubits'
distance is far enough so that the direct coupling between them
could be neglected. Then we can concentrate on discussing the role
of the bath in this model. First, we use Holstein-Primakoff
transformation to reduce the model to a effective ``spin-boson''
Hamiltonian. Then we apply a numerical simulation to obtain the
reduced dynamics of the two-spin qubits. During our numerical
calculation, there are no approximations assumed and the initial
state of the subsystem (consisted by the two spin qubits) can be
arbitrary. It is well-known that the concurrence is a measure of
entanglement degree between two spin qubits and the fidelity is
also an important property, that has been widely applied into
quantum coding theory \cite{Schumacher}. Thus some results about
these quantities in the thermal limit will be given in the latter
part of this paper. The rest of this paper is organized as
following. In Sec. \ref{Hamiltonian} the model Hamiltonian and the
operator transformation procedure is introduced. In Sec.
\ref{calculation}, we explain the numerical techniques about the
evolution of the reduced matrix for the subsystem. Detailed
results and discussions can be found in Sec. \ref{discussion}. We
will conclude our study in Sec. \ref{conclusion}.

\section{Model and Transformation}\label{Hamiltonian}

Consider a two-spin-qubit subsystem symmetrically interacting with
bath spins via a Heisenberg XY interaction: both the subsystem and
the bath are composed of spin-1/2 atoms. Every spin in the bath
interacts with each of the two center spins of equal strength,
similar to the cases considered in \cite{Breuer, Yuan, Hutton,
Breuer3}. The Hamiltonian for the total system is divided as three
parts:
\begin{eqnarray}\label{Hamip}
H&=&H_S+H_{SB}+H_B.\\ \label{H_S}
H_S&=&\mu_0(\sigma_{01}^z+\sigma_{02}^z),\\ \label{HSBp}
H_{SB}&=&\frac{g_0}{2\sqrt{N}}\sum_{i=1}^N\left[(1+\gamma)
(\sigma_{01}^x\sigma_i^x+\sigma_{02}^x\sigma_i^x)+
(1-\gamma)(\sigma_{01}^y\sigma_i^y+\sigma_{02}^y\sigma_i^y)\right],\\
\label{HB} H_B&=&\frac{g}{2N}\sum_{i\neq
j}^N\left[(1+\gamma)\sigma_i^x\sigma_j^x+
(1-\gamma)\sigma_i^y\sigma_j^y\right].
\end{eqnarray}
Here, $H_S$ and $H_B$ are the Hamiltonians of the subsystem and
bath respectively, and $H_{SB}$ describes the interaction between
them \cite{Breuer, Yuan, Canosa}. $\mu_0$ represents the coupling
constant between a locally applied external magnetic field in the
$z$ direction and the spin qubit subsystem. $\gamma$,
$-1\leq\gamma\leq1$ is the anisotropic parameter. When $\gamma=0$,
it is of an isotropic case \cite{Yuan}. In the following part of
this paper, we only talk about cases with positive $\gamma$ for
the symmetry of the spin star structure. $\sigma_{0i}^x$,
$\sigma_{0i}^y$ and $\sigma_{0i}^z$ ($i$=1,2) are the operators of
the qubit subsystem spins, respectively. By Pauli matrix, the
operators read
\begin{equation}
\sigma^x=\left(\begin{array}{cc}
      0 & 1 \\
      1 & 0
    \end{array}\right), \quad
\sigma^y=\left(\begin{array}{cc}
      0 & -i\\
      i & 0
    \end{array}\right), \quad
\sigma^z=\left(\begin{array}{cc}
      1 & 0\\
      0 & -1
    \end{array}\right).
\end{equation}
$\sigma_i^x$ and $\sigma_i^y$ are the corresponding operators of the
$i$th atom spin in the bath. The indices $i$ of the summation for
the spin bath run from $1$ to $N$, where $N$ is the number of the
bath atoms. $g_0$ is the coupling constant between the qubit
subsystem spins and bath spins, whereas $g$ is the coupling between
the bath spins.\\

Using $\sigma^x=(\sigma^++\sigma^-),
\sigma^y=-i(\sigma^+-\sigma^-)$, we can rewrite Hamiltonians.
(\ref{HSBp}) and (\ref{HB}) as:
\begin{equation}\label{HSBp1}
\begin{split}
H_{SB}=&\frac{g_0}{\sqrt{N}}\biggl[\sum_{i=1}^N\sigma_i^+(\gamma\sigma_{01}^++
\sigma_{01}^-)+\sum_{i=1}^N\sigma_i^-(\sigma_{01}^++
\gamma\sigma_{01}^-)\\   &+
\sum_{i=1}^N\sigma_i^+(\gamma\sigma_{02}^++
\sigma_{02}^-)+\sum_{i=1}^N\sigma_i^-(\sigma_{02}^++
\gamma\sigma_{02}^-)\biggr],
\end{split}
\end{equation}
\begin{equation}\label{HBp1}
H_B=\frac{g}{N}\sum_{i\neq
j}^N\left[\gamma(\sigma_i^+\sigma_j^++\sigma_i^-\sigma_j^-)+
(\sigma_i^+\sigma_j^-+\sigma_i^-\sigma_j^+)\right].
\end{equation}

Substituting the collective angular momentum operators
$J_{\pm}=\sum_{i=1}^N\sigma_{i}^\pm$ into Eqs. (\ref{HSBp1}) and
(\ref{HBp1}), we get
\begin{eqnarray}\label{H_SB1}
H_{SB}&=&\frac{g_0}{\sqrt{2j}}\left[J_+(\gamma\sigma_{01}^++
\sigma_{01}^-)+J_-(\sigma_{01}^++\gamma\sigma_{01}^-)+
J_+(\gamma\sigma_{02}^++\sigma_{02}^-)+J_-(\sigma_{02}^++
\gamma\sigma_{02}^-)\right],\\ \label{H_B1}
H_B&=&\frac{g}{2j}\left[\gamma\left(J_+J_++J_-J_-\right)+
\left(J_+J_-+J_-J_+-2j\right)\right].
\end{eqnarray}
where $j=N/2$. After the Holstein-Primakoff transformation
\cite{Holstein},
\begin{equation}\label{J}
J_+=b^+(\sqrt{2j-b^+b}), \hspace*{2mm} J_-=(\sqrt{2j-b^+b})b,
\end{equation}
with $[b,b^{+}]=1$ and in the thermodynamic limit (i.e.
$N\longrightarrow\infty$) at finite temperatures, the Hamiltonian,
Eqs. (\ref{H_SB1}) and (\ref{H_B1}), can finally be written as
\begin{eqnarray}\label{H_SB}
H_{SB}&=&g_0\left[b^+(\gamma\sigma_{01}^++
\sigma_{01}^-+\gamma\sigma_{02}^++\sigma_{02}^-)+b(\sigma_{01}^++\gamma\sigma_{01}^-+
\sigma_{02}^++\gamma\sigma_{02}^-)\right],\\ \label{H_B}
H_B&=&g[\gamma({b^+}^2+b^2)+ 2b^+b].
\end{eqnarray}

The transformed Hamiltonian describes two qubits interacting with a
single-mode thermal bosonic bath field, so the analysis of the model
is just like a nontrivial problem in the field of cavity quantum
electrodynamics \cite{Imamoglu, Zheng}. We note here that due to the
transition invariance of the bath spins in our model, it is
effectively represented by a single collective environment
pseudo-spin $J$ in Eq. (\ref{J}). After the Holstein-Primakoff
transformation and in the thermodynamic limit, this collective
environment pseudo-spin could be considered a single-mode bosonic
thermal field. The effect of this single-mode environment on the
dynamics of the two qubits is interesting. In Sec. \ref{discussion},
we will show some results, for example, the revival behavior of the
reduced density matrix or entanglement evolution of the subsystem
spins. This can be used in real quantum information application.

\section{Numerical Calculation procedures}\label{calculation}

The initial density matrix of the total system is assumed to be
separable, i.e., $\rho(0)=|\psi\rangle\langle\psi|\otimes\rho_B$.
The density matrix of the spin bath satisfies the Boltzmann
distribution, that is $\rho_B=e^{-H_B/T}/Z$, where $Z={\rm
Tr}\left(e^{-H_B/T}\right)$ is the partition function, and the
Boltzmann constant $k_B$ has been set to $1$ for simplicity. The
density matrix $\rho(t)$ of the whole system can formally be derived
by
\begin{eqnarray}
\rho(t)&=&\exp(-iHt)\rho(0)\exp(iHt),\\ \label{rhoB}
\rho(0)&=&\rho_S(0)\otimes\rho_B(0),\\ \label{rho_s}
\rho_S(0)&=&|\psi(0)\rangle\langle\psi(0)|.
\end{eqnarray}
In order to find the density matrix $\rho(t)$, we follow the method
suggested by Tessieri et al. \cite{TWmodel}. The thermal bath state
$\rho_B(0)$ can be expanded with the eigenstates of the environment
Hamiltonian $H_B$ in Eq. (\ref{H_B}):
\begin{eqnarray}\label{rho_B}
\rho_B(0)&=&\sum_{m=1}^M|\phi_m\rangle\omega_m\langle\phi_m|,\\
\label{weight}
\omega_m&=&\frac{e^{-E_m/T}}{Z},\\
Z&=&\sum_{m=1}^Me^{-E_m/T}.
\end{eqnarray}
Here $|\phi_m\rangle$, $m=1, 2, 3, \cdots, M$, are the eigenstates
of $H_B$, and $E_m$ the corresponding eigenenergies in increasing
order. $M$ is just the number of eigenstates considered in this
summation. With this expansion, the density matrix $\rho(t)$ can
be written as:
\begin{equation}\label{equ:2m}
\rho(t)=\sum_{m=1}^M\omega_m|\Psi_m(t)\rangle\langle\Psi_m(t)|.
\end{equation}
Where
\begin{equation}
|\Psi_m(t)\rangle =\exp(-iHt)|\Psi_m(0)\rangle
=U(t)|\Psi_m(0)\rangle.
\end{equation}
The initial state is
\[
|\Psi_m(0)\rangle =|\psi(0)\rangle|\phi_m\rangle.
\]
The evolution operator $U(t)$ can be evaluated by different methods.
In Ref. \cite{Yuan}, they use a unique analytical operator
technique. Here, we apply an efficient numerical algorithm based on
polynomial schemes \cite{Jing, Dobrovitski1, Hu} into this problem.
The method used in this calculation is the Laguerre polynomial
expansion method we proposed in Ref. \cite{Jing}, which is pretty
well suited to many quantum systems, open or closed, and can give
accurate result in a much smaller computation load. More precisely,
the evolution operator $U(t)$ is expanded in terms of the Laguerre
polynomial of the Hamiltonian as:
\begin{equation*}
U(t)=\left(\frac{1}{1+it}\right)^{\alpha+1}
\sum^{\infty}_{k=0}\left(\frac{it}{1+it}\right)^kL^{\alpha}_k(H).
\end{equation*}
$L^{\alpha}_k(H)$ is one type of Laguerre polynomials \cite{Arfken}
as a function of $H$, where $\alpha$ ($-1<\alpha<\infty$)
distinguishes different types of the Laguerre polynomials and $k$ is
the order of it. In real calculations the expansion has to be cut at
some value of $k_{\text{max}}$, which was optimized to be $20$ in
this study (We have to test out a $k_{\text{max}}$ for the
compromise of the numerical stability in the recurrence of the
Laguerre polynomial and the speed of calculation). With the largest
order of the expansion fixed, the time step $t$ is restricted to
some value in order to get accurate results of the evolution
operator. At every time step, the accuracy of the results will be
confirmed by the test of the numerical stability --- whether the
trace of the density matrix is $1$ with error less than $10^{-12}$.
For longer times the evolution can be achieved by more steps. The
action of the Laguerre polynomial of Hamiltonian to the states is
calculated by recurrence relations of the Laguerre polynomial. The
efficiency of this polynomial scheme \cite{Jing} is about $9$ times
as that of the Runge-Kutta algorithm under the same accuracy
condition used in Ref. \cite{TWmodel}. When the states
$|\Psi_m(t)\rangle$ are obtained, the density matrix can be obtained
by performing a summation in Eq. (\ref{equ:2m}).\\

Although theoretically we should consider every energy state of
the single-mode bath field: $M\rightarrow\infty$, but the
contributions of the high energy states $|\phi_m\rangle, m>m_C$
($m_C$ is a cutoff to the spin bath eigenstates) are found to be
neglectable due to their very tiny weight value $\omega_m$, as
long as the temperature is finite. That is to say, the $M$ in Eqs.
(\ref{rho_B}) to (\ref{equ:2m}) could be changed to $m_C$. Then we
use the following equation in real calculation:
\begin{equation}\label{equ:M}
\rho(t)=\sum_{m=1}^{m_C}\omega_m|\Psi_m(t)\rangle\langle\Psi_m(t)|.
\end{equation}

After obtaining the density matrix of the whole system, the reduced
density matrix is calculated by a partial trace operation to
$\rho(t)$, which trace out the degrees of freedom of the
environment:
\begin{equation}\label{final}
\rho_S(t)={\rm Tr}_B\left(\rho(t)\right).
\end{equation}
For the model of this paper, $\rho_S=|\psi\rangle\langle\psi|$ is
the density matrix of the open subsystem consists of two separate
spins, which can be expressed as a $4\times4$ matrix in the Hilbert
space of the subsystem spanned by the orthonormal vectors
$|00\rangle$, $|01\rangle$, $|10\rangle$ and $|11\rangle$. The most
general form of an initial pure state of the two-qubit system is
\begin{eqnarray}
|\psi(0)\rangle=\alpha|00\rangle+\beta|11\rangle+\gamma|01\rangle+\delta|10\rangle,\\
\mbox{with}\quad |\alpha|^2+|\beta|^2+|\gamma|^2+|\delta|^2=1.
\end{eqnarray}

\section{Numerical simulation results and discussions}
\label{discussion}

When the reduced density matrix is determined, any physical
quantities of the subsystem can be readily found out. In the
following we will discuss three important physical quantities of
the subsystem which reflect the decoherence speed, the
entanglement degree and the fidelity of the subsystem state. These
quantities are (i) the moment of spin-01, here we choose the first
spin $\langle\sigma^{z}_{01}\rangle$, which demonstrates the
decoherence rate of the system; (ii) the concurrence
\cite{Wootters1, Wootters2} for the two spins of the open
subsystem. The concurrence of the two spin-1/2 system is an
indicator of their intra entanglement, which is defined as
\cite{Wootters1}:
\begin{equation}\label{Concurrence}
C=\max\{\lambda_1-\lambda_2-\lambda_3-\lambda_4,~0\},
\end{equation}
where $\lambda_i$ are the square roots of the eigenvalues of the
product matrix
$\rho_S(\sigma^y\otimes\sigma^y)\rho^*_S(\sigma^y\otimes\sigma^y)$
in decreasing order; (iii) the fidelity \cite{Privman}, which is
defined as
\begin{equation}\label{fide}
Fd(t)={\rm Tr}_S[\rho_{\rm ideal}(t)\rho(t)].
\end{equation}
$\rho_{\rm ideal}(t)$ represents the pure-state evolution of the
subsystem under $H_S$ only, without interaction with the
environment. The fidelity is a measure for decoherence and depends
on $\rho_{ideal}$, is equal to one only if the time dependent
density matrix $\rho(t)$ is equal to $\rho_{ideal}(t)$. The
corresponding results and discussions are divided to two subsections
according to different initial states. \\

For the product states, the present paper focuses on the
entanglement generation by the spin bath and dose not involve the
revival of the initial state for $C(t=0)=0$. Thus in the subsection
\ref{product}, we give out the dynamics of concurrence and
$\sigma_{01}^{z}$. For the Bell states, since the system can evolve
to a completely different state from the initial one and has the
same concurrence $C(t)>0$, we should give out the evolution of
concurrence and fidelity in subsection \ref{bell}.

\subsection{Product states}\label{product}

\begin{figure}[htbp]
\centering \subfigure[$C(t)$]{\label{I11ga:C}
\includegraphics[width=3in]{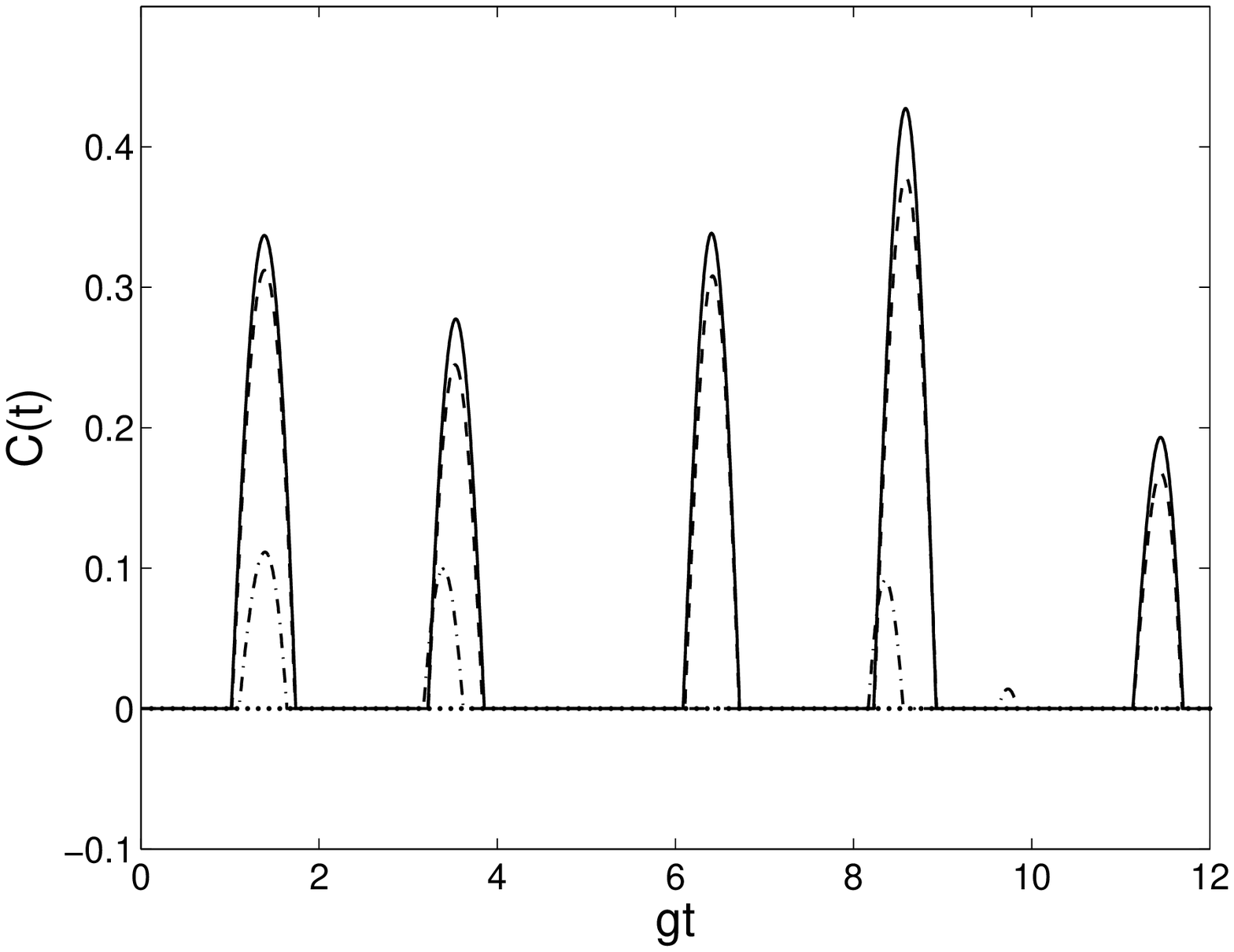}}
\subfigure[$\langle\sigma^{z}_{01}(t)\rangle$]{\label{I11ga:Sz}
\includegraphics[width=3in]{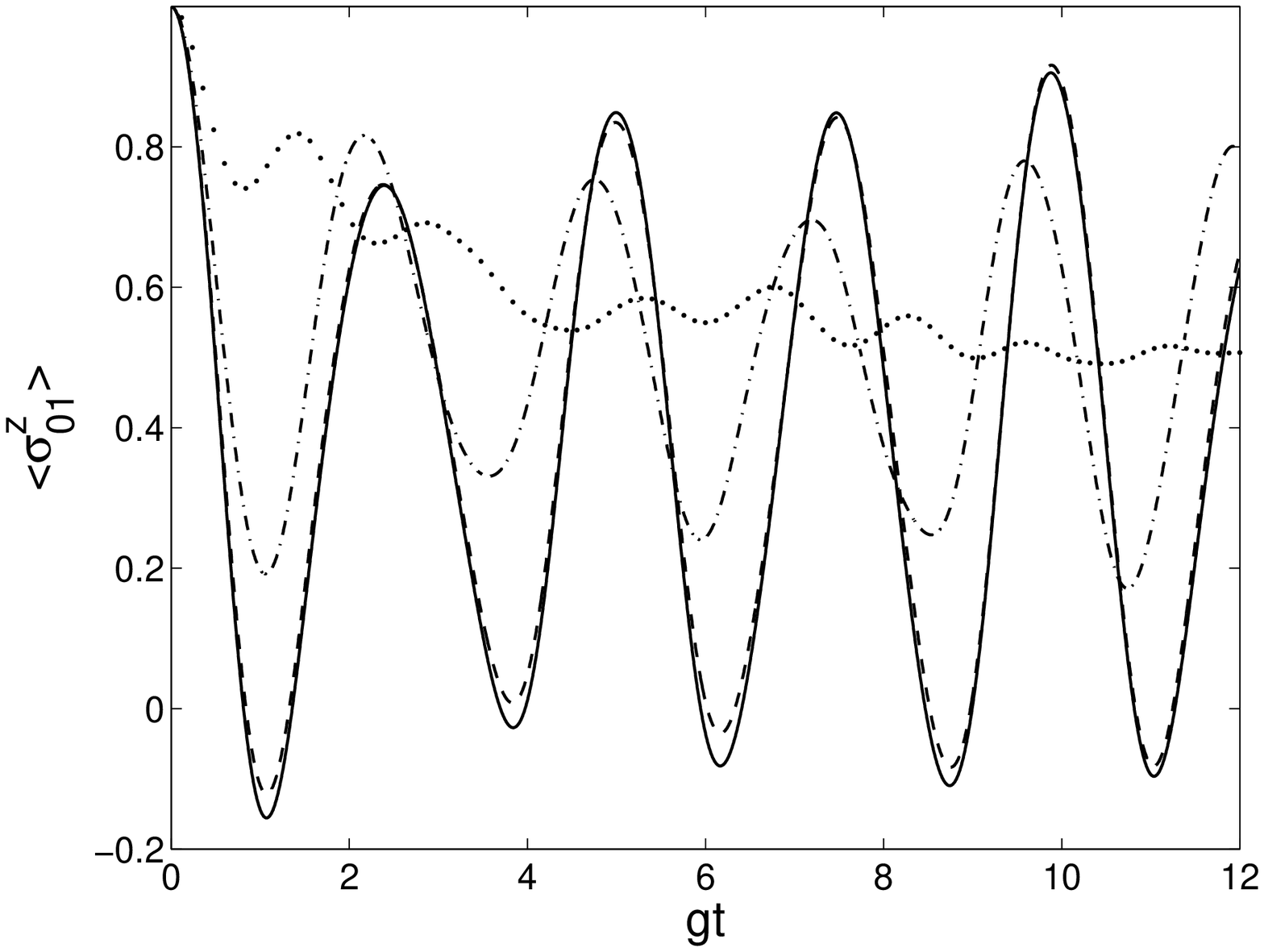}}
\caption{Time evolution for (a) Concurrence, (b) the moment of
spin-01 from an initial two-qubit state of
$|\psi(0)\rangle=|11\rangle$ at different values of anisotropic
parameter: $\gamma=0$ (solid curve), $\gamma=0.2$ (dashed curve),
$\gamma=0.6$ (dot dashed curve), $\gamma=1.0$ (dotted curve).
Other parameters are $\mu_0=2g$, $g_0=g$, $T=g$.} \label{I11ga}
\end{figure}

\begin{figure}[htbp]
\centering \subfigure[$C(t)$]{\label{I11T:C}
\includegraphics[width=3in]{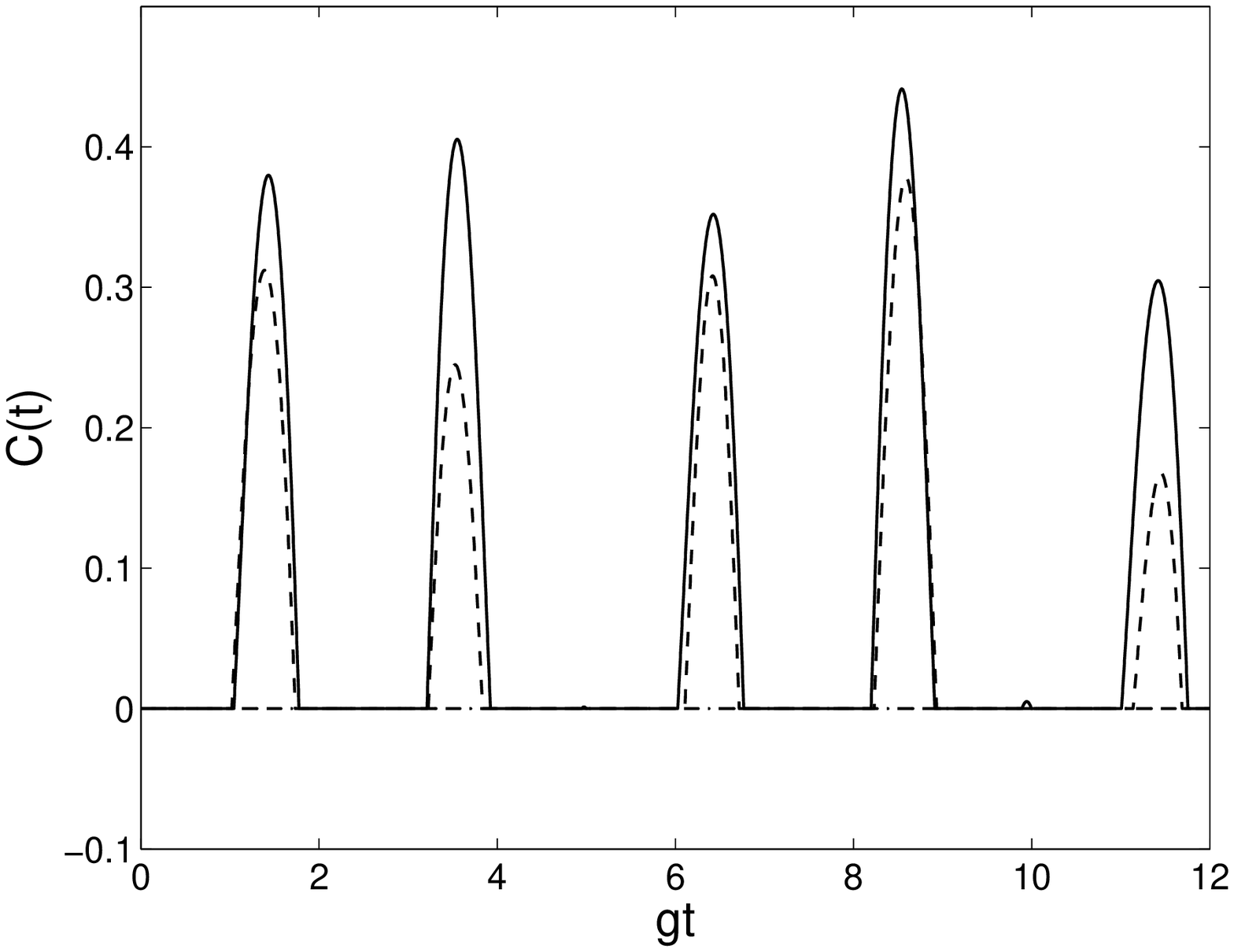}}
\subfigure[$\langle\sigma^{z}_{01}(t)\rangle$]{\label{I11T:Sz}
\includegraphics[width=3in]{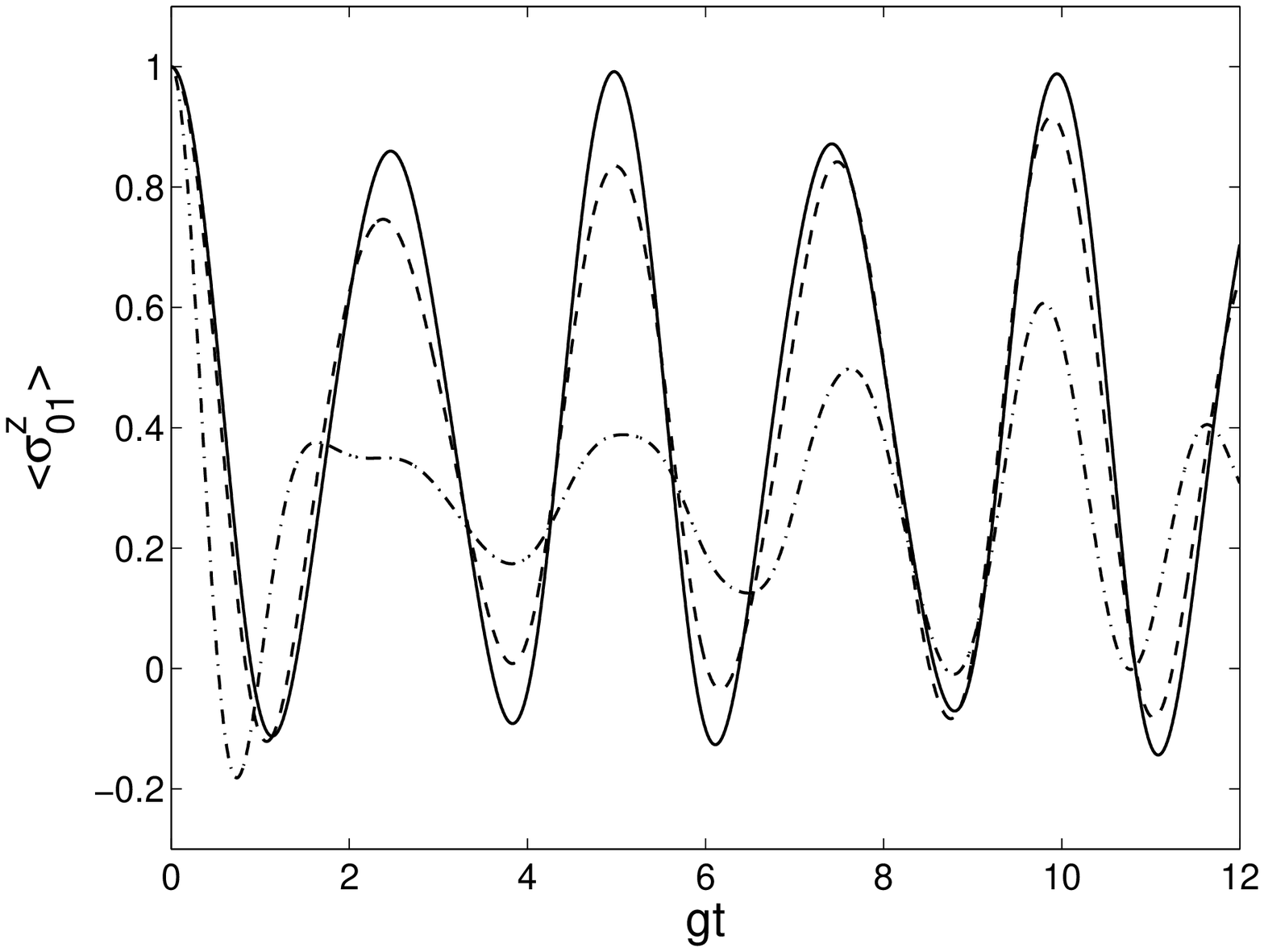}}
\caption{Time evolution for (a) Concurrence, (b) the moment of
spin-01 from an initial two-qubit state of
$|\psi(0)\rangle=|11\rangle$ at different values of temperature:
$T=0.2g$ (solid curve), $T=g$ (dashed curve), $T=5g$ (dot dashed
curve). Other parameters are $\mu_0=2g$, $g_0=g$, $\gamma=0.2$.}
\label{I11T}
\end{figure}

\begin{figure}[htbp]
\centering \subfigure[$C(t)$]{\label{I01ga:C}
\includegraphics[width=3in]{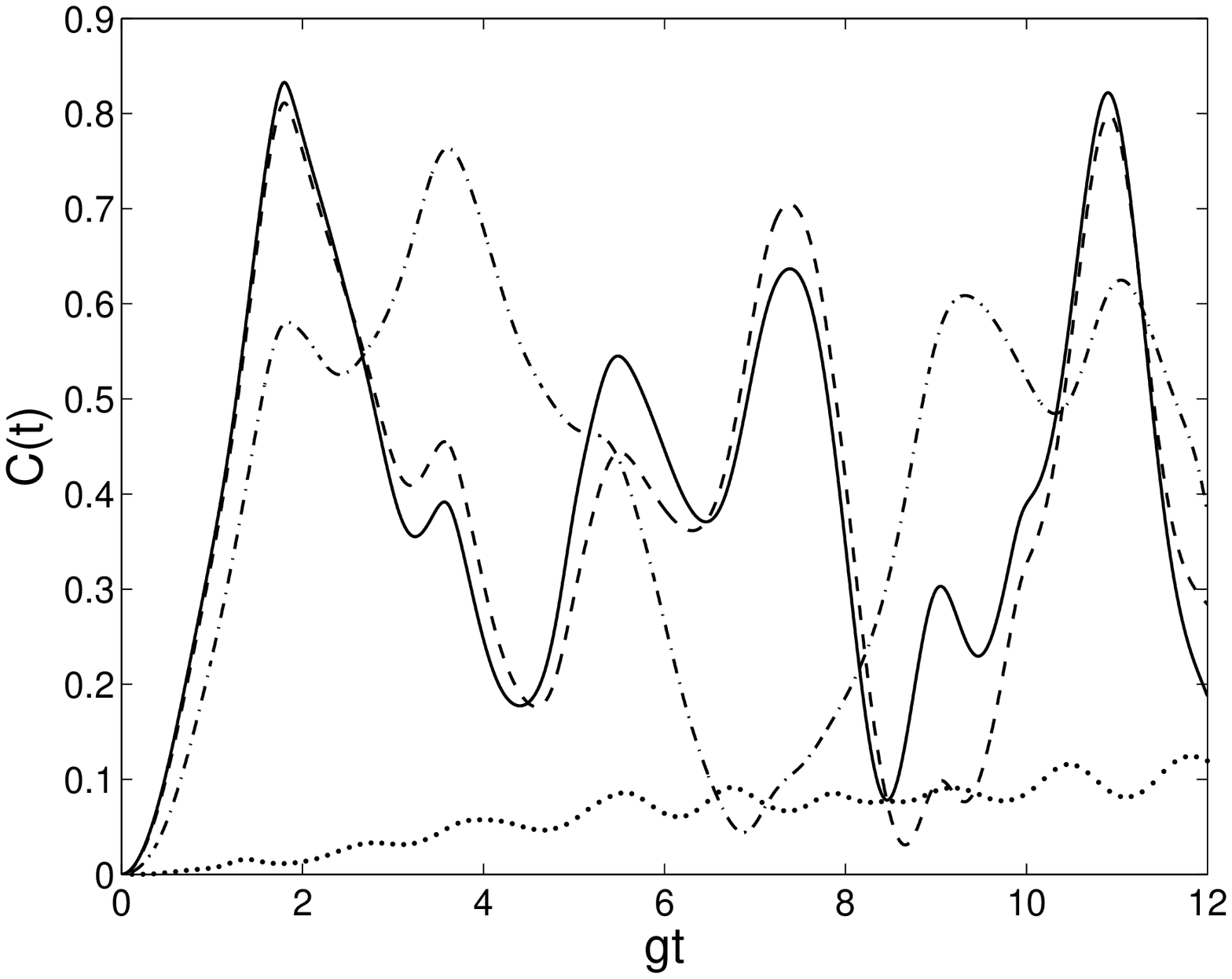}}
\subfigure[$\langle\sigma^{z}_{01}(t)\rangle$]{\label{I01ga:Sz}
\includegraphics[width=3in]{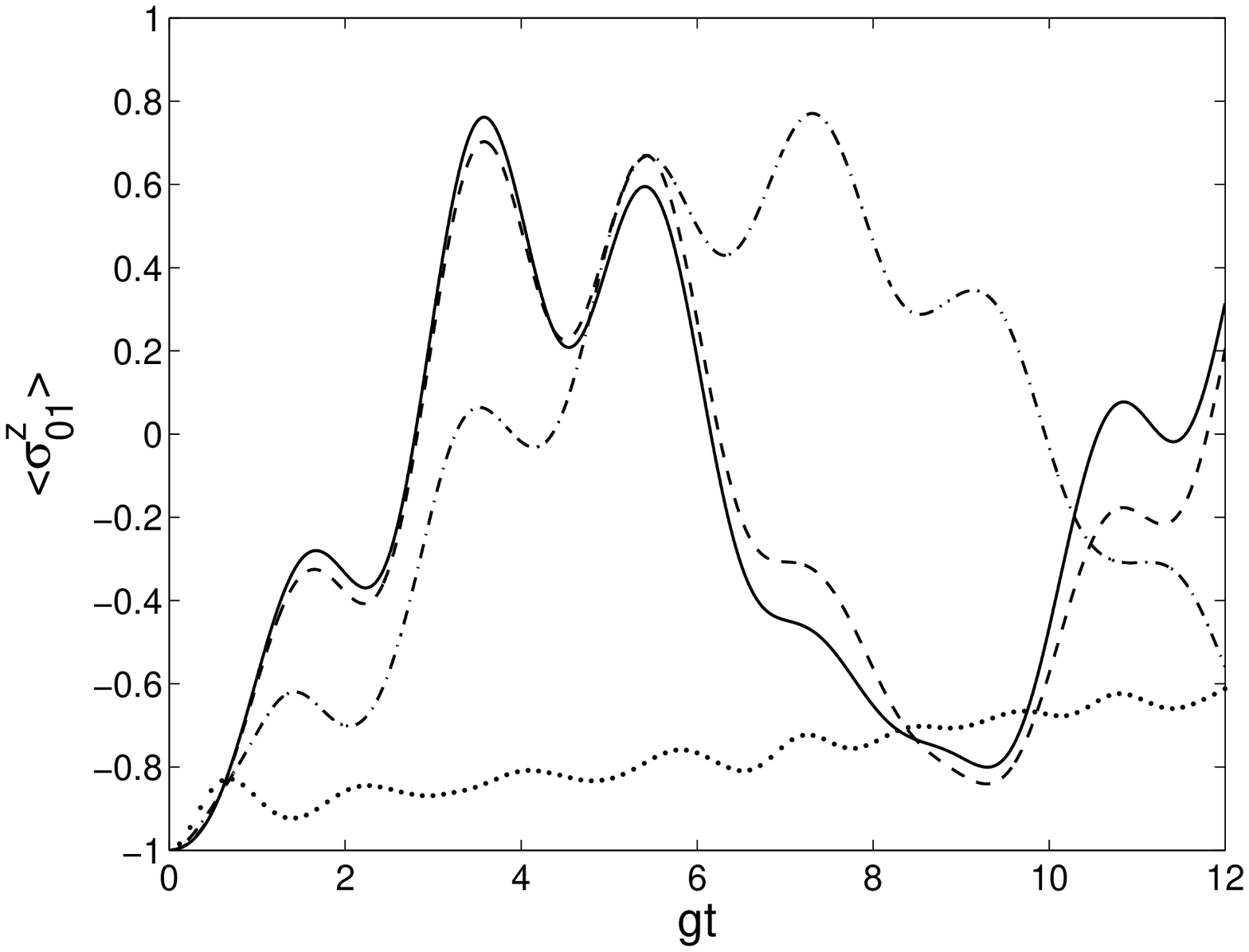}}
\caption{Time evolution for (a) Concurrence, (b) the moment of
spin-01 from an initial two-qubit state of
$|\psi(0)\rangle=|01\rangle$ at different values of anisotropic
parameter: $\gamma=0$ (solid curve), $\gamma=0.2$ (dashed curve),
$\gamma=0.6$ (dot dashed curve), $\gamma=1.0$ (dotted curve).
Other parameters are $\mu_0=2g$, $g_0=g$, $T=g$.} \label{I01ga}
\end{figure}

\begin{figure}[htbp]
\centering \subfigure[$C(t)$]{\label{I01T:C}
\includegraphics[width=3in]{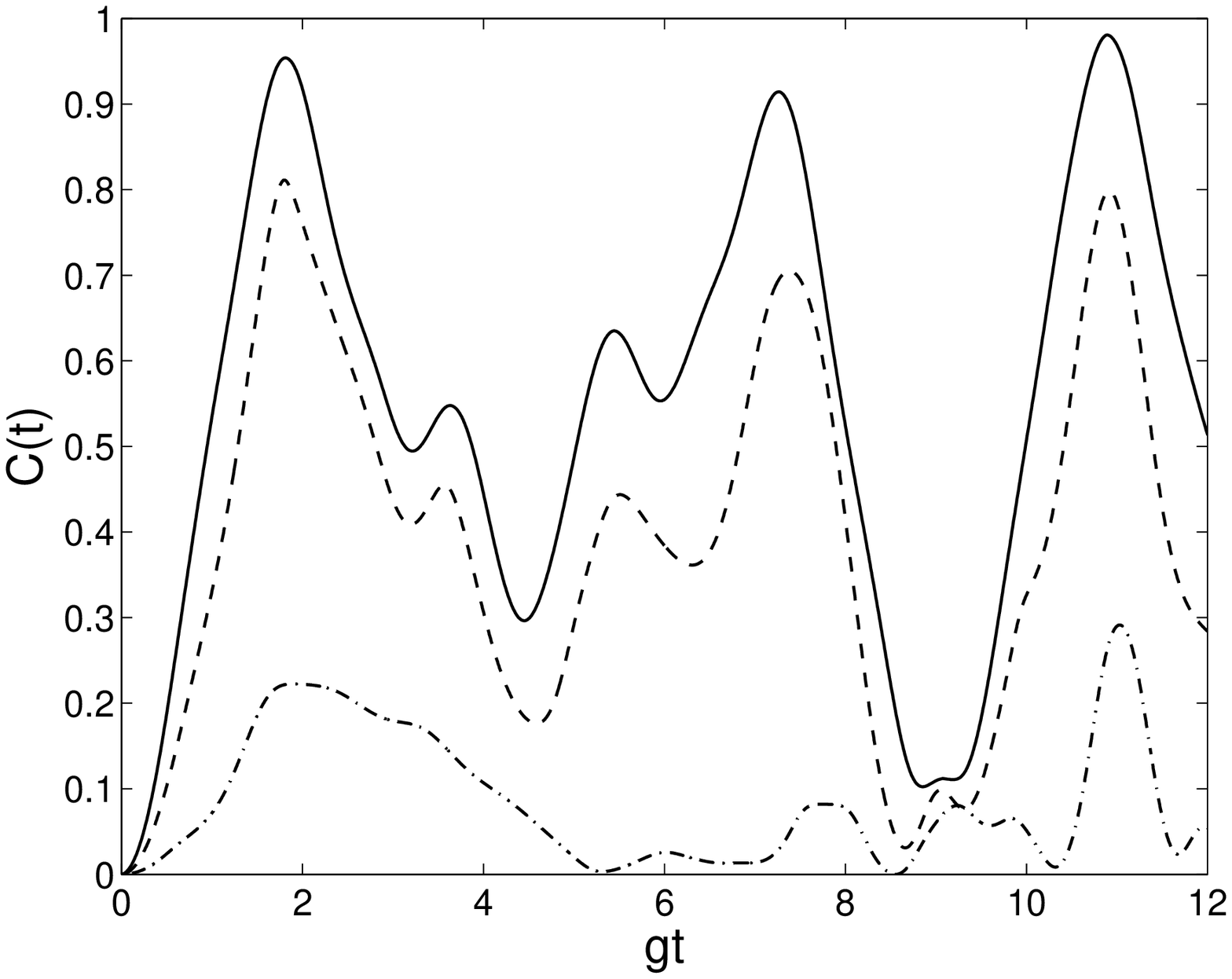}}
\subfigure[$\langle\sigma^{z}_{01}(t)\rangle$]{\label{I01T:Sz}
\includegraphics[width=3in]{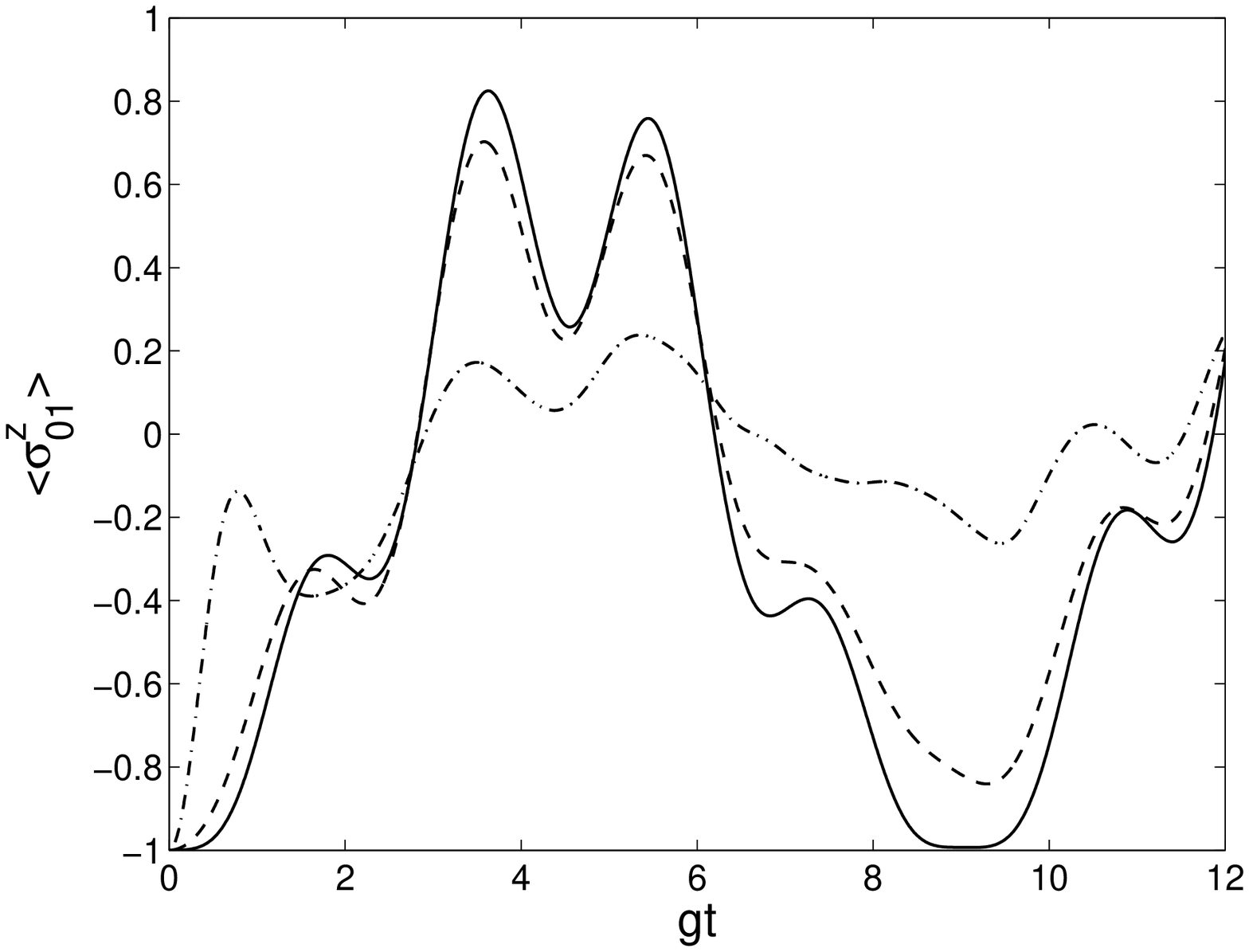}}
\caption{Time evolution for (a) Concurrence, (b) the moment of
spin-01 from an initial two-qubit state of
$|\psi(0)\rangle=|01\rangle$ at different values of temperature:
$T=0.2g$ (solid curve), $T=g$ (dashed curve), $T=5g$ (dot dashed
curve). Other parameters are $\mu_0=2g$, $g_0=g$, $\gamma=0.2$.}
\label{I01T}
\end{figure}

\begin{figure}[htbp]
\centering \subfigure[$C(t)$]{\label{I01g0:C}
\includegraphics[width=3in]{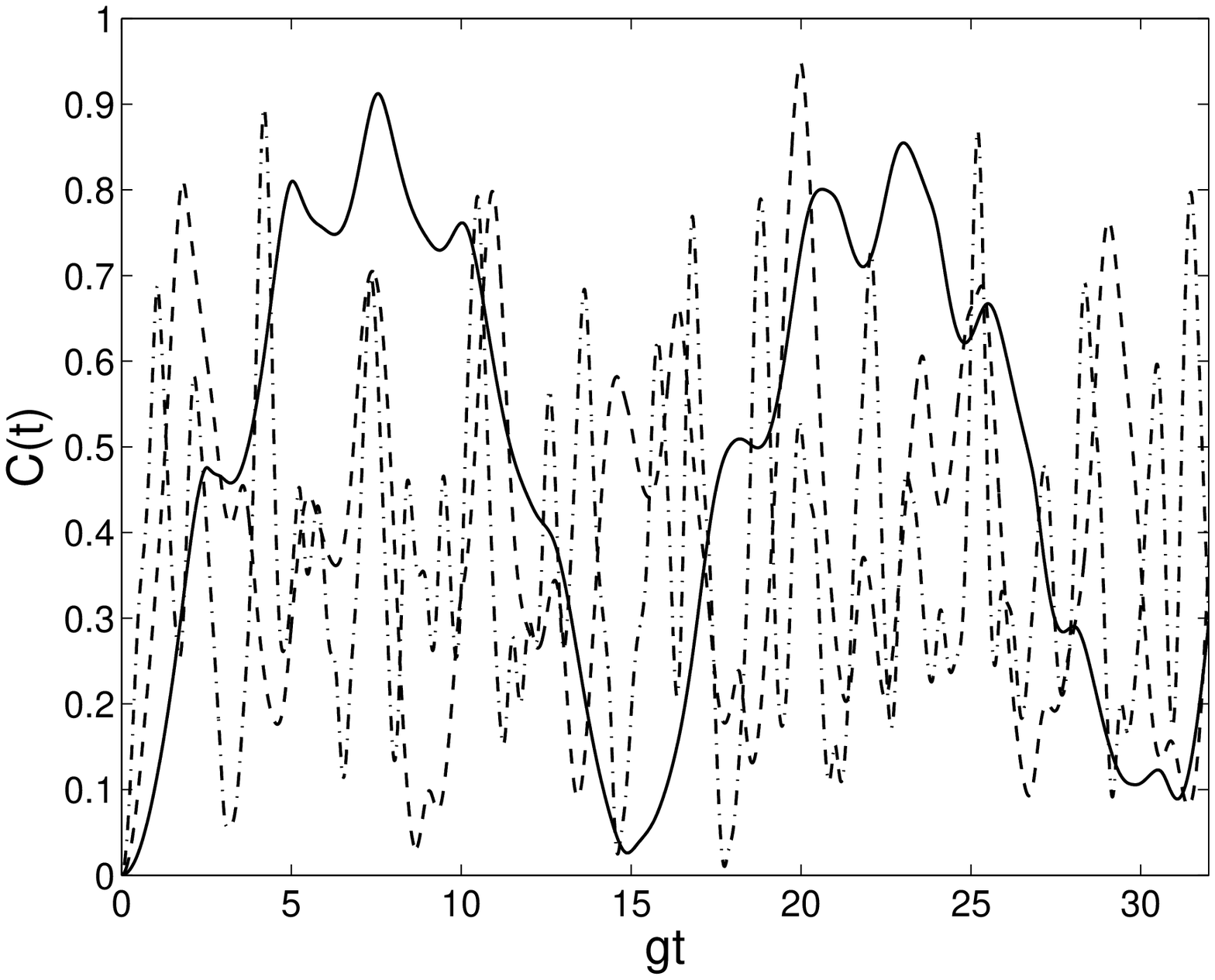}}
\subfigure[$\langle\sigma^{z}_{01}(t)\rangle$]{\label{I01g0:Sz}
\includegraphics[width=3in]{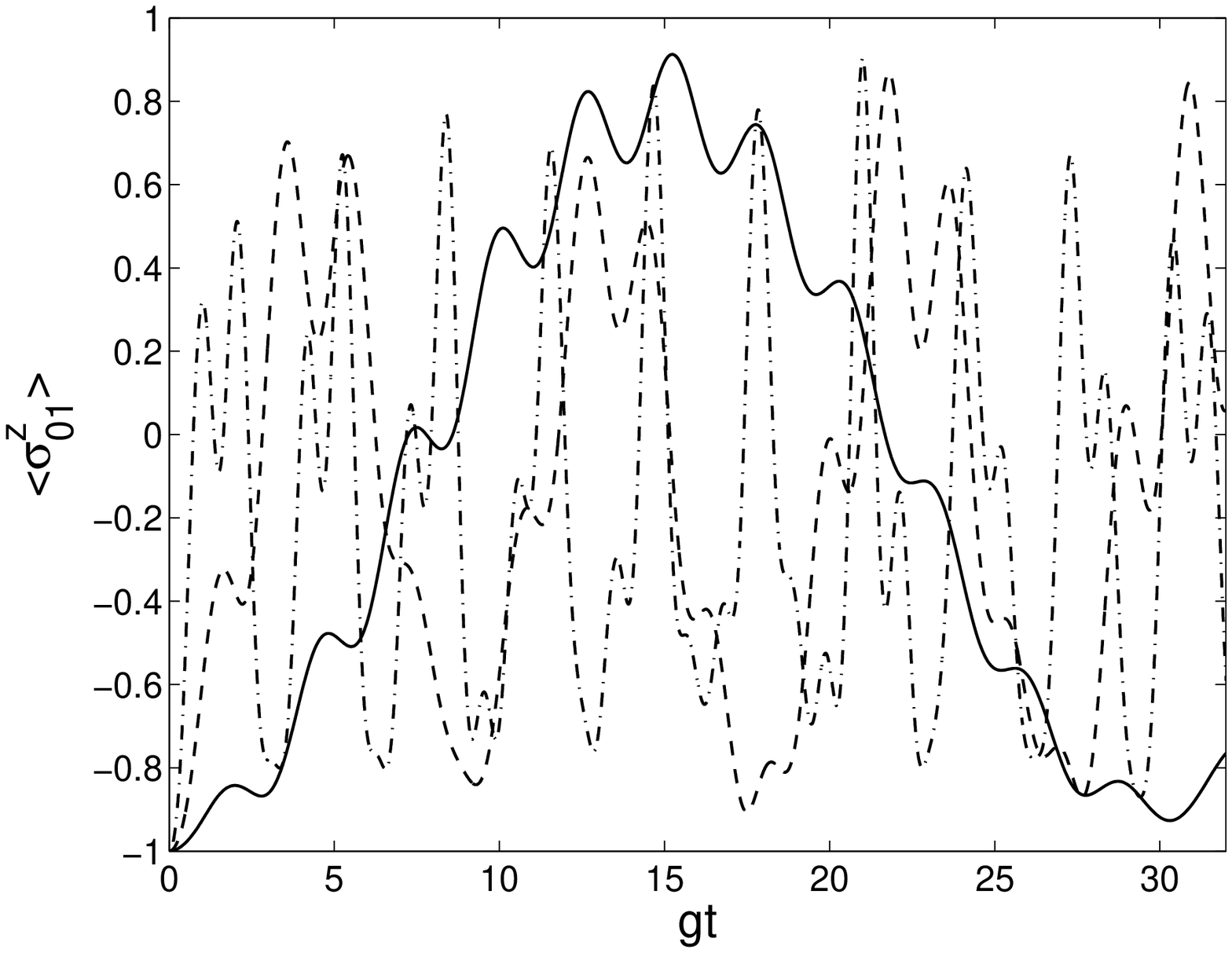}}
\caption{Time evolution for (a) Concurrence, (b) the moment of
spin-01 from an initial two-qubit state of
$|\psi(0)\rangle=|01\rangle$ at different values of coupling
strength between subsystem and bath: $g_0=0.5g$ (solid curve),
$g_0=g$ (dashed curve), $g_0=2g$ (dot dashed curve). Other
parameters are $\mu_0=2g$, $\gamma=0.2$, $T=g$.} \label{I01g0}
\end{figure}

First we show the evolution of concurrence and $z$ component moment
as functions of anisotropic parameter $\gamma$ from two initial
product states $|11\rangle$ (to see Fig. \ref{I11ga}) and
$|10\rangle$ (to see Fig. \ref{I01ga}). It is obvious that the
entanglement between the two subsystem qubits can arise from the
interaction with the bath. Yet this kind of effect of the spin-bath
is decreased with increasing the anisotropic parameter $\gamma$. And
the above variation depends on the initial states: with
$|\psi(0)\rangle=|11\rangle$ (Fig. \ref{I11ga:C}), when
$\gamma>0.87$, the concurrence of the two qubits will be always kept
zero as initialed; with $\psi(0)=|01\rangle$ (Fig. \ref{I01ga:C}),
the entanglement can always be created to some extend; and if
$\gamma$ approaches to $0$ (the isotropic case), the concurrence can
increase as high as $0.8$ over some period of oscillations. In Fig.
\ref{I11ga:Sz} and Fig. \ref{I01ga:Sz}, with $\gamma$ increasing,
the oscillation amplitudes of the curves become smaller and smaller,
which means that the coherence of the subsystem approaches to be
lost. For the initial state $|11\rangle$, after the first spin flip
(the sign of $\sigma_{01}^z$ changes from positive to negative) for
$\gamma\geq0.6$, it can not flip again. However, for
$|\psi(0)\rangle=|10\rangle$, the spin can flip after a period of
time even for $\gamma=0.6$. Therefore, the increase of entanglement
depends sensitively on the anisotropic parameter. \\

The bath is in a thermal equilibrium state at different temperature,
which effect is shown in Fig. \ref{I11T} and Fig. \ref{I01T}. In
these two figures, the anisotropic parameter $\gamma$ is kept as
$0.2$. We can find that (i) at a very low temperature,
$\sigma^{z}_{01}(t)$ displays a nearly periodical oscillation, which
is identical with the two-photon resonance of two two-level atoms in
a cavity. And the subsystem entanglement can be raised to a
comparatively degree; (ii) with increasing temperature, the
oscillation amplitudes of the curves are damped due to the thermal
bath. For the concurrence, $C(t)\rightarrow0$ means to approach a
most separated state (to see the dot dashed curve in Fig.
\ref{I11T:C} and Fig. \ref{I01T:C}). For $\sigma_{01}^z$, it means
the degeneration of its magnetic moment (to see the dot dashed curve
in Fig. \ref{I11T:Sz} and Fig. \ref{I01T:Sz}). Therefore it is clear
that the subsystem loses its memory faster as the temperature
increases. \\

In Fig. \ref{I11ga:C} and \ref{I01ga:C}, we can find that the
entanglement between the two initial separated spins can be
generated with the help of the single-mode thermal bosonic bath
field. Assume that the system is initially prepared in $|01\rangle$
(or $|10\rangle$), a pure state. On one hand, when the interaction
between the system and spin bath is turned on, one spin could drop
from the excited state and simultaneously the other spin could jump
absorbing the boson just emitted by the former. This process induces
the entanglement of two spins. On the other hand, it also evolves
into a mixed state resorting to the bosons provided by the
single-mode boson field.  Thus, the coupling between the system and
its environment leads to the entanglement between two initial
separated qubits.

Then we keep the bath at a moderate temperature $T=1g$ to find out
the effect of the coupling strength $g_0$ in Fig. \ref{I01g0}. At
a smaller value $g_0=0.2g$, the weak interaction with the bath
will make both the concurrence and $\sigma_{01}^z$ display a
pseudo-periodical behavior; on the contrary, at a larger value
$g_0=5g$, their dynamics is too strongly disturbed by the bath to
be utilized. Thus in real applications, the coupling between the
subsystem and the spin bath should be reduced.

\subsection{Bell states}\label{bell}

\begin{figure}[htbp]
\centering \subfigure[$C(t)$]{\label{I1001ga:C}
\includegraphics[width=3in]{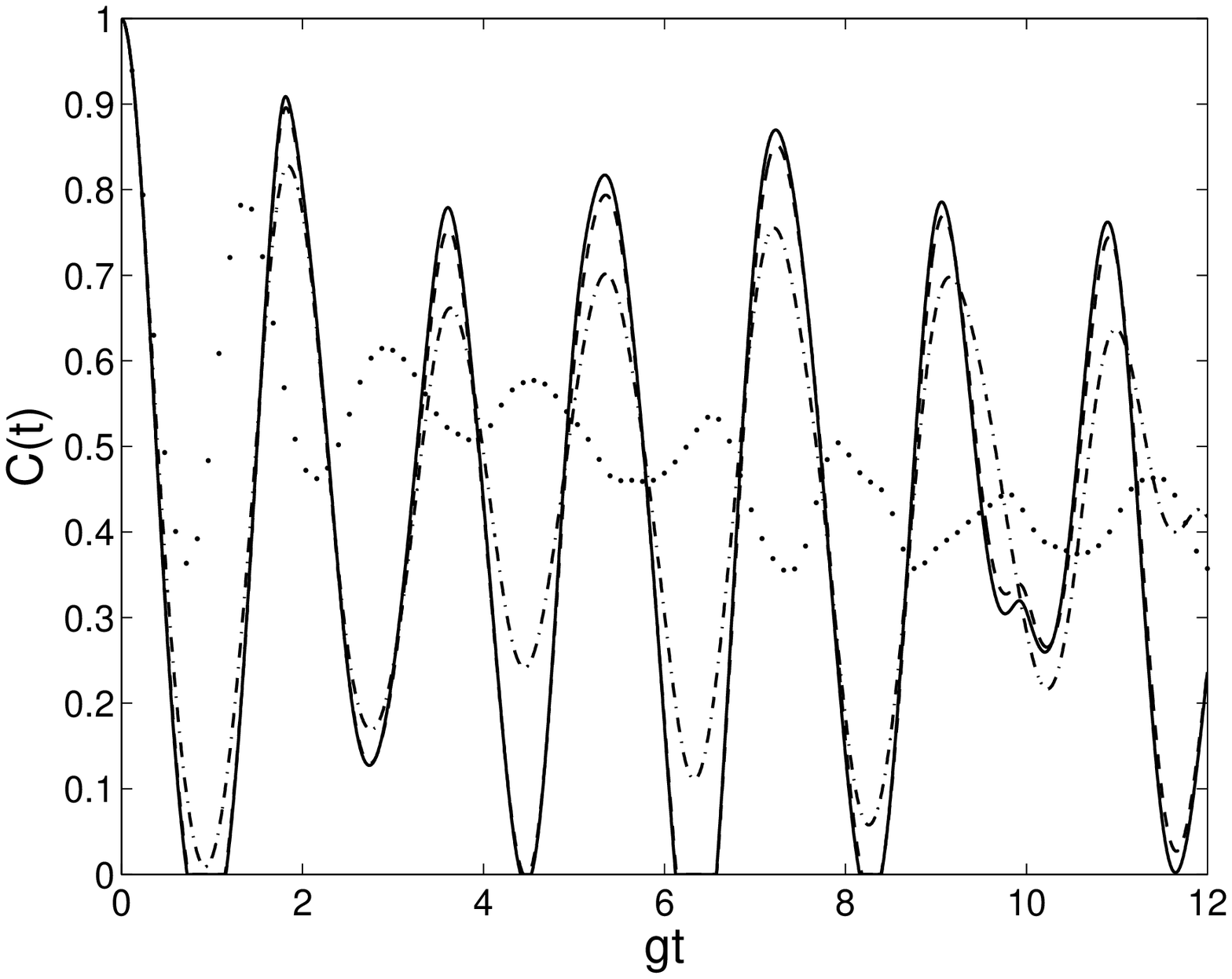}}
\subfigure[$Fd(t)$]{\label{I1001ga:Fd}
\includegraphics[width=3in]{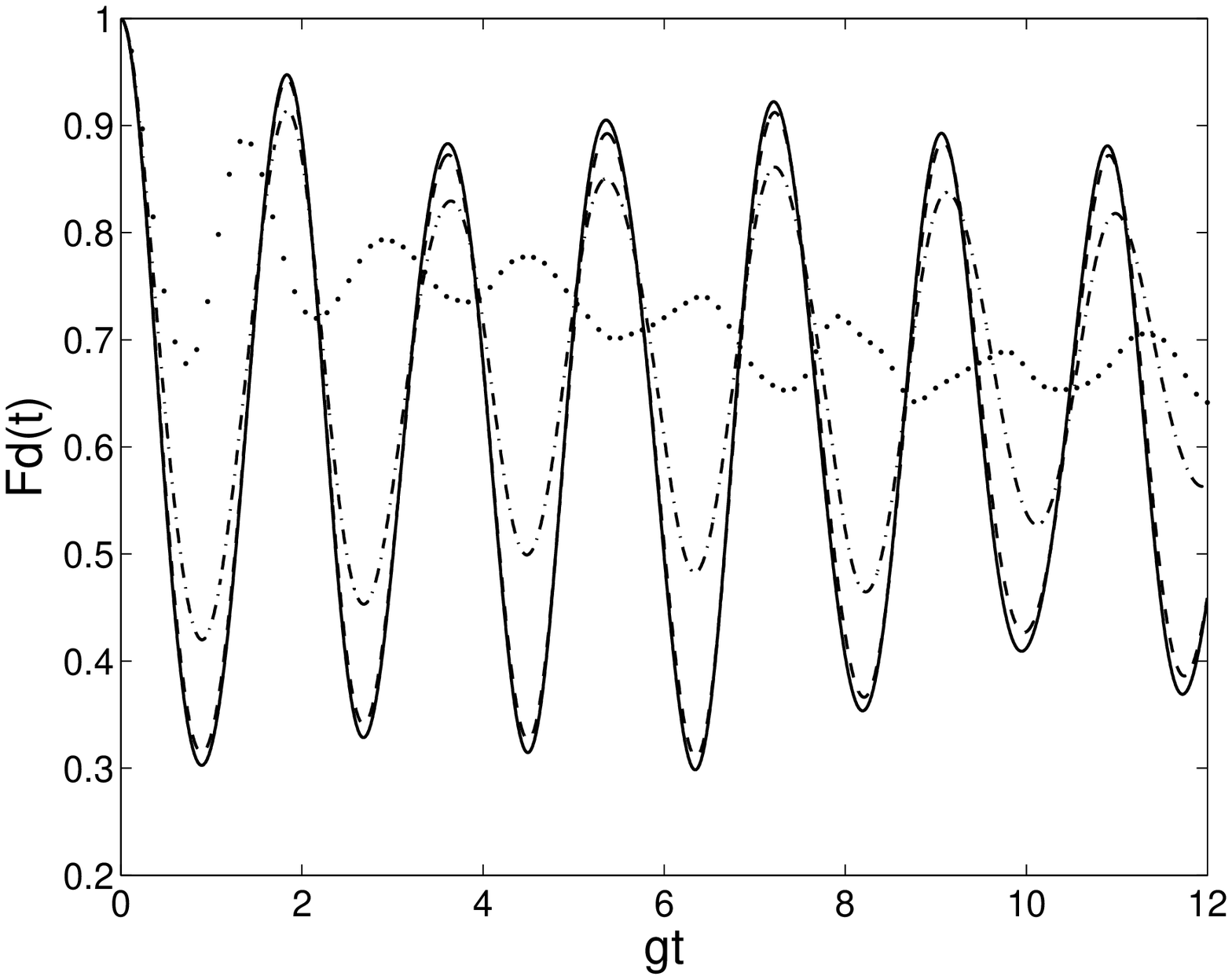}}
\caption{Time evolution for (a) Concurrence, (b) Fidelity from an
initial two-qubit state of
$|\psi(0)\rangle=1/\sqrt{2}(|10\rangle+|01\rangle)$ at different
values of anisotropic parameter: $\gamma=0$ (solid curve),
$\gamma=0.2$ (dashed curve), $\gamma=0.6$ (dot dashed curve),
$\gamma=1.0$ (dotted curve). Other parameters are $\mu_0=2g$,
$g_0=g$, $T=g$.} \label{I1001ga}
\end{figure}

\begin{figure}[htbp]
\centering \subfigure[$C(t)$]{\label{I1001T:C}
\includegraphics[width=3in]{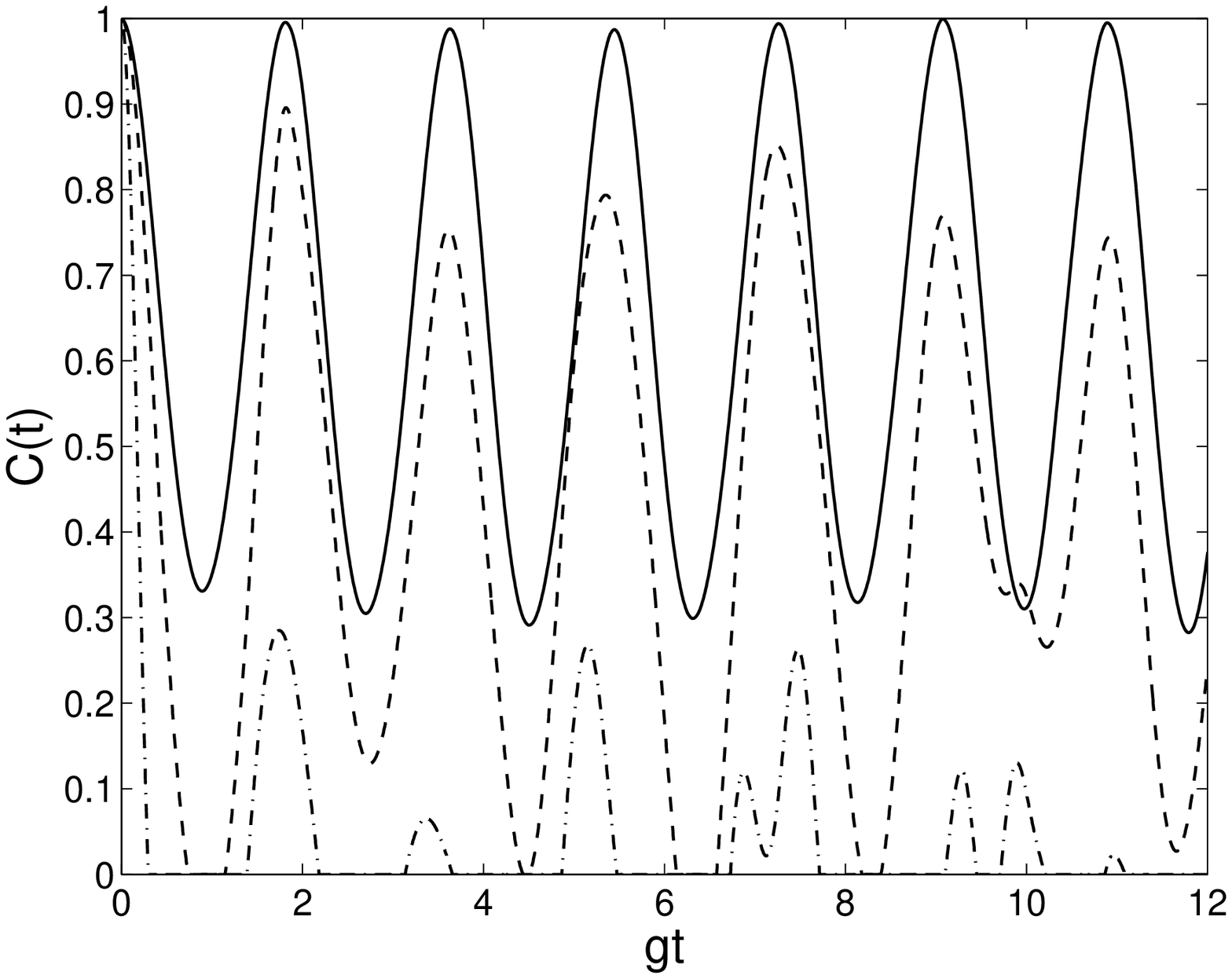}}
\subfigure[$Fd(t)$]{\label{I1001T:Fd}
\includegraphics[width=3in]{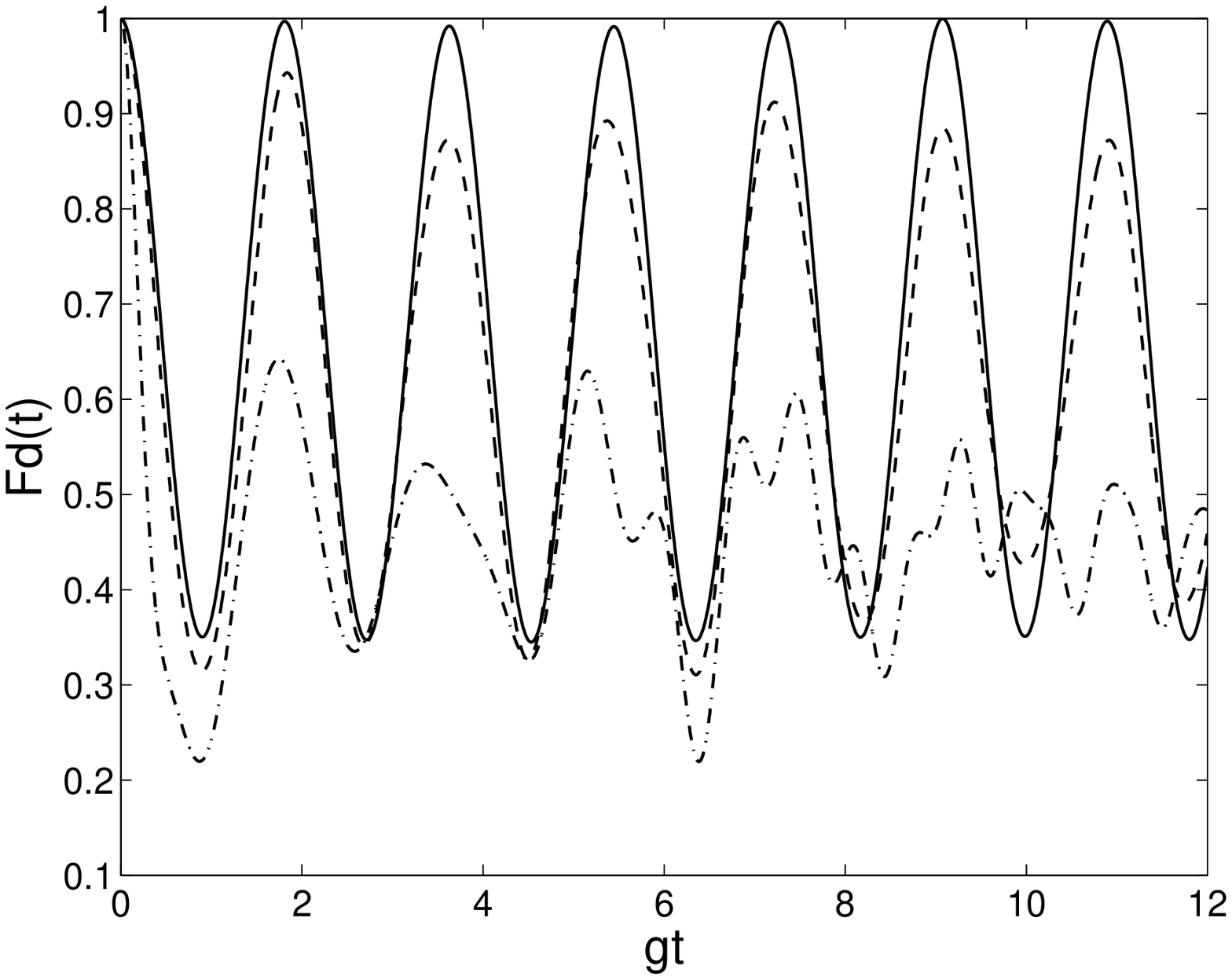}}
\caption{Time evolution for (a) Concurrence, (b) Fidelity from an
initial two-qubit state of
$|\psi(0)\rangle=1/\sqrt{2}(|10\rangle+|01\rangle)$ at different
values of temperature: $T=0.2g$ (solid curve), $T=g$ (dashed
curve), $T=5g$ (dot dashed curve). Other parameters are
$\mu_0=2g$, $g_0=g$, $\gamma=0.2$.} \label{I1001T}
\end{figure}

\begin{figure}[htbp]
\centering \subfigure[$C(t)$]{\label{I1100ga:C}
\includegraphics[width=3in]{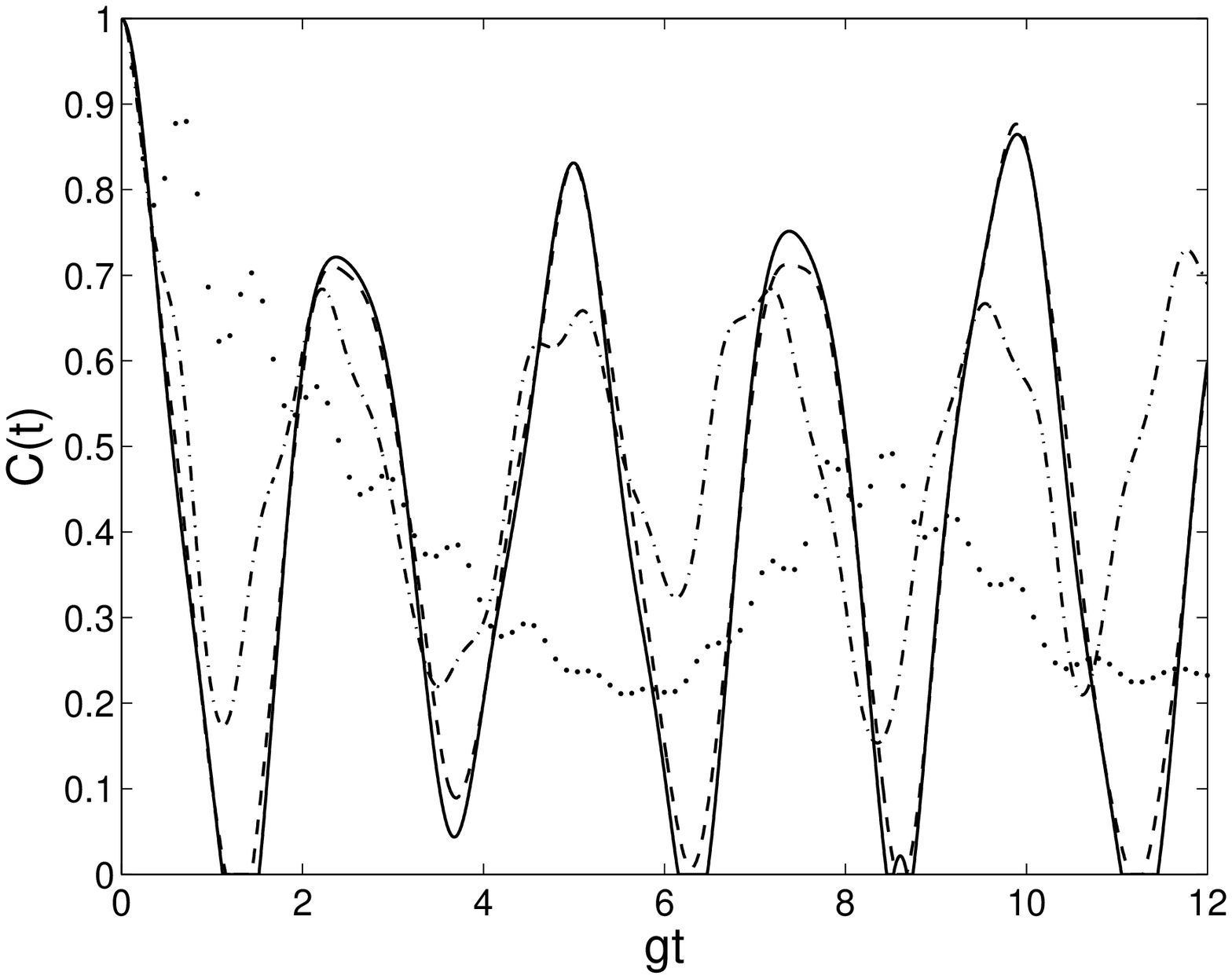}}
\subfigure[$Fd(t)$]{\label{I1100ga:Fd}
\includegraphics[width=3in]{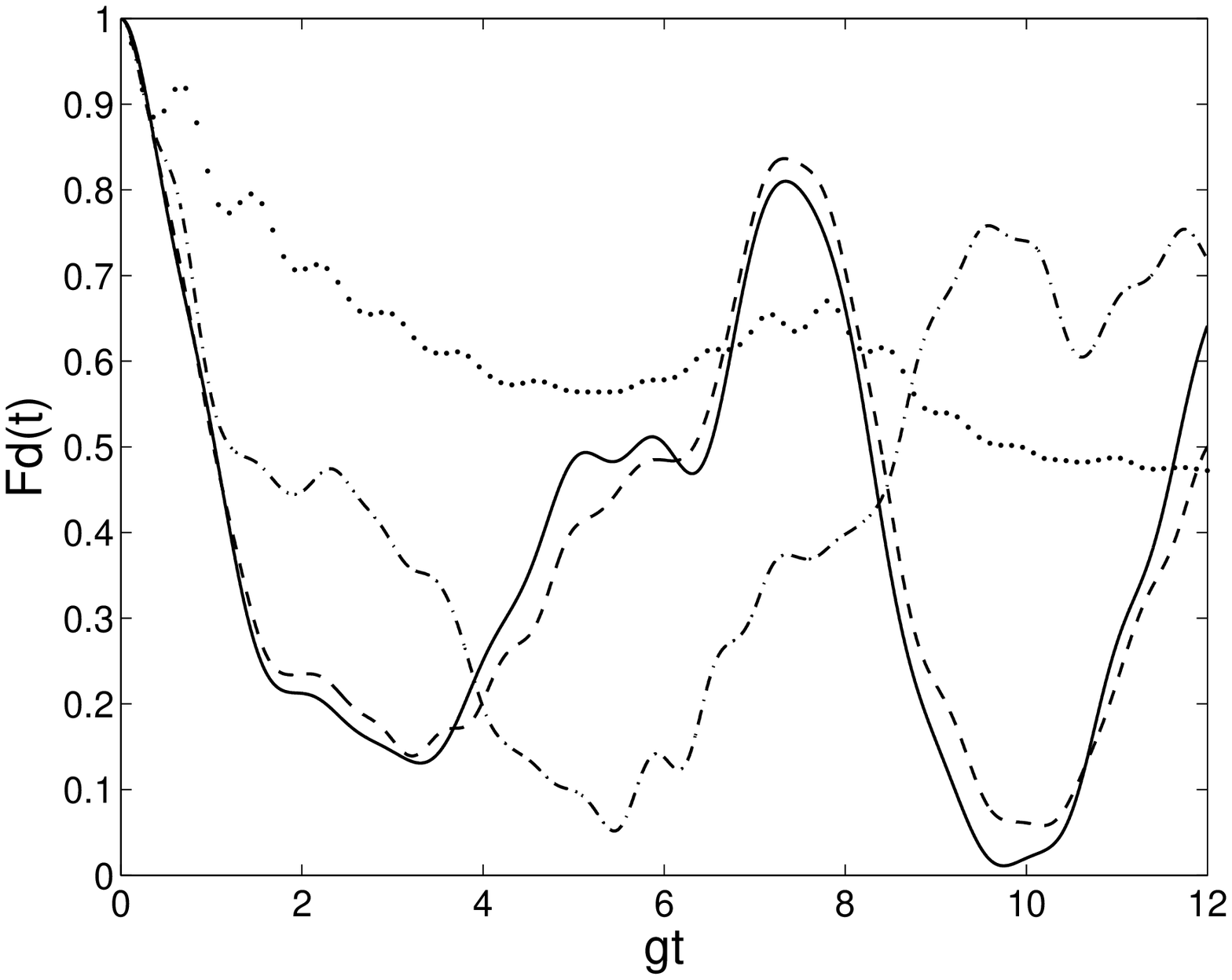}}
\caption{Time evolution for (a) Concurrence, (b) Fidelity from an
initial two-qubit state of
$|\psi(0)\rangle=1/\sqrt{2}(|11\rangle+|00\rangle)$ at different
values of anisotropic parameter: $\gamma=0$ (solid curve),
$\gamma=0.2$ (dashed curve), $\gamma=0.6$ (dot dashed curve),
$\gamma=1.0$ (dotted curve). Other parameters are $\mu_0=2g$,
$g_0=g$, $T=g$.} \label{I1100ga}
\end{figure}

\begin{figure}[htbp]
\centering \subfigure[$C(t)$]{\label{I1100T:C}
\includegraphics[width=3in]{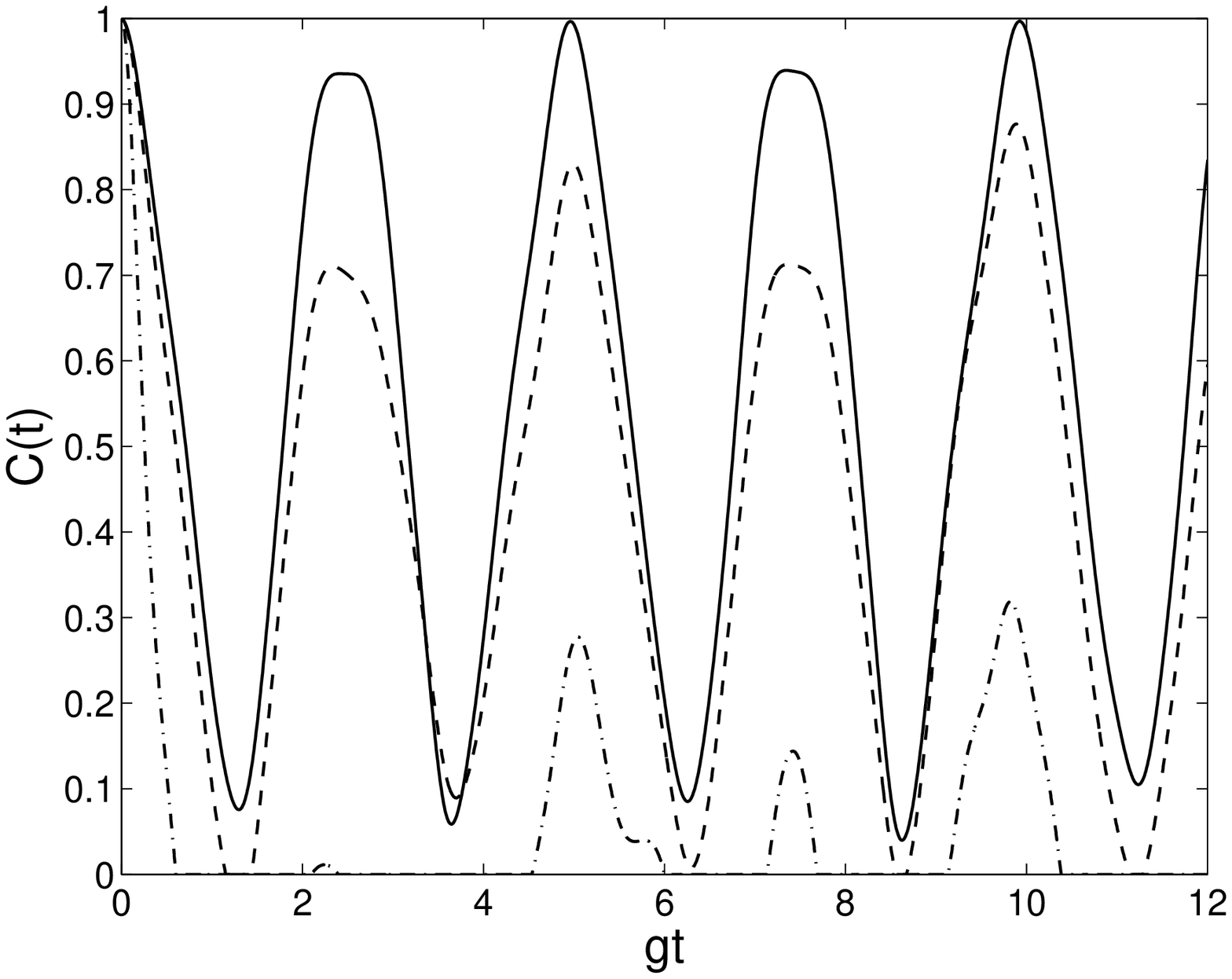}}
\subfigure[$Fd(t)$]{\label{I1100T:Fd}
\includegraphics[width=3in]{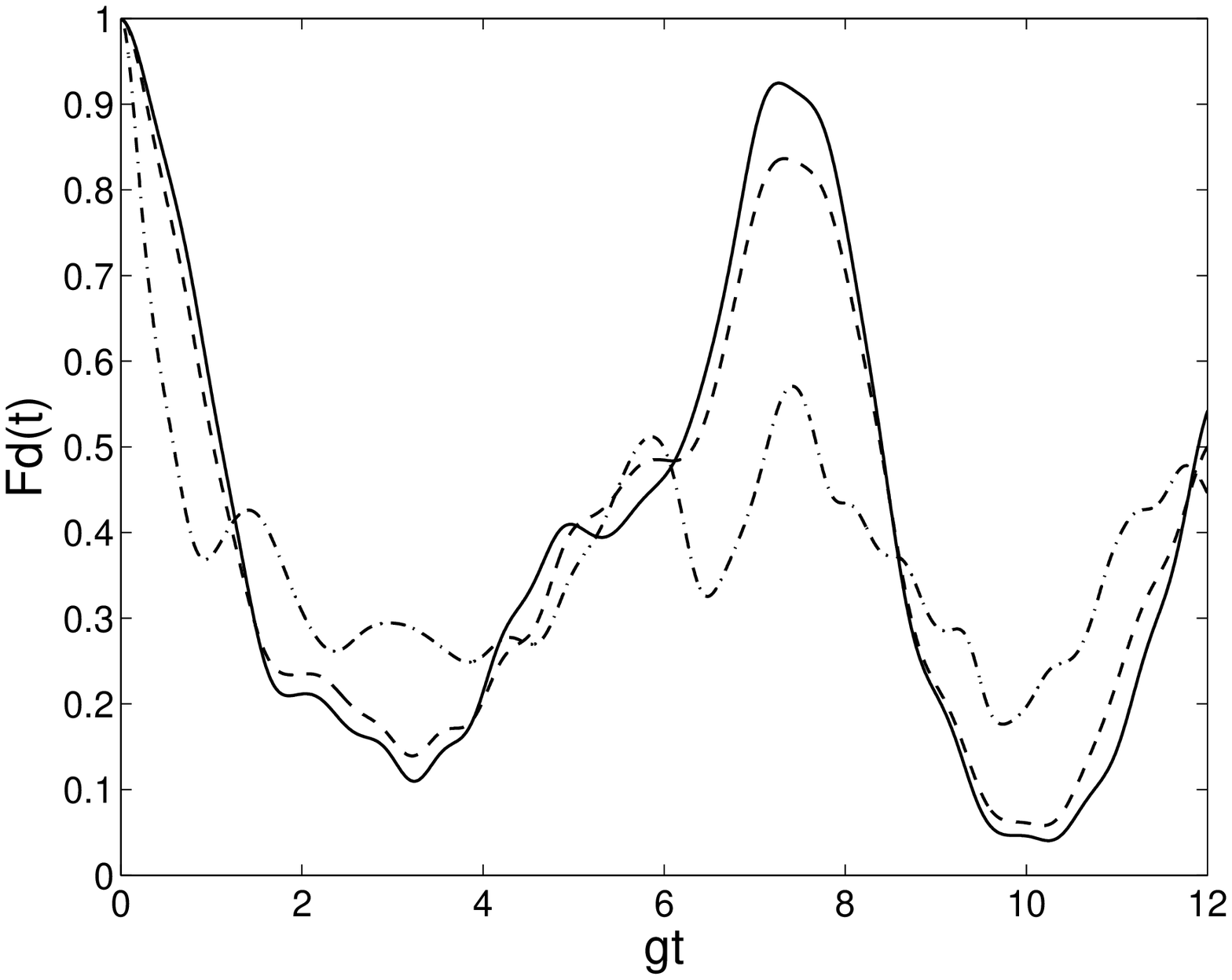}}
\caption{Time evolution for (a) Concurrence, (b) Fidelity from an
initial two-qubit state of
$|\psi(0)\rangle=1/\sqrt{2}(|11\rangle+|00\rangle)$ at different
values of temperature: $T=0.2g$ (solid curve), $T=g$ (dashed curve),
$T=5g$ (dot dashed curve). Other parameters are $\mu_0=2g$, $g_0=g$,
$\gamma=0.2$.} \label{I1100T}
\end{figure}

In the cases that the subsystem is prepared as a most entangled
state $C=1$ (Bell states), the anisotropic parameter $\gamma$
still makes an important effect on the time evolutions of
concurrence and the $\sigma_{01}^z$. When
$\psi(0)=1/\sqrt{2}(|10\rangle+|01\rangle)$, the concurrence (to
see Fig. \ref{I1001ga:C}) of subsystem is always revived to
$C\approx0.8$ after some time of oscillation at small value of
$\gamma$. The results can be proved by the revival fidelity of the
subsystem in Fig. \ref{I1001ga:Fd}: at every summit, the state
mainly consists of its initial state. But an interesting phenomena
is found by the comparison of Fig. \ref{I1100ga:C} with Fig.
\ref{I1100ga:Fd}. It is noticed that two summits disappear in the
interval of $0.0<gt<8.0$! Thus we analyze the states at the first
three summits in Fig. \ref{I1100ga:C}. It is found that at
$gt=2.448$, the most component of the subsystem state is
$1/\sqrt{2}(|11\rangle-|00\rangle)$; at $gt=4.960$, the state of
the two qubits is very near to a combination of
$1/\sqrt{2}(|11\rangle-i|00\rangle)$ and
$1/\sqrt{2}(|11\rangle-|00\rangle)$; at $gt=7.480$, the main part
of the state comes back to its initial state. So the concurrence
can not determine the concrete state of the subsystem in the
present case. Even if the concurrence can be restored, the state
is not always the same as the initial one. Only the combination of
the concurrence and the fidelity gives the information of real
state evolution.\\

In Fig. \ref{I1001T} and Fig. \ref{I1100T}, we plot the dynamics
behavior of the concurrence and fidelity at different temperatures.
When the temperature is as low as $T=0.2g$, both cases of
$\psi(0)=1/\sqrt{2}(|10\rangle+|01\rangle)$ and
$\psi(0)=1/\sqrt{2}(|11\rangle+|00\rangle)$ display a periodical
oscillation, the concurrence can always nearly restore its initial
value. But for the former case, the dynamics of fidelity is
synchronous with that of concurrence; for the latter, the revival of
the concurrence does not always mean that of the state. From the
viewpoint of the definition of fidelity (\ref{fide}), it is partly
due to the system part of the Hamiltonian (\ref{H_S}). In the
special case of the bell state $1/\sqrt{2}(|10\rangle+|01\rangle)$,
$\rho_{ideal}(t)$ is identical to the spins initial density matrix
(It is an eigenstate of $H_S$), while for
$\psi(0)=1/\sqrt{2}(|11\rangle+|00\rangle)$ $\rho_{ideal}(t)$ is not
in the same condition. It is further proved that the properties of
the dynamics should be determined by the combination of concurrence
and fidelity.

\section{Conclusion}\label{conclusion}

We have studied the dynamics evolution of two separated qubit spins
in a bath consisted of infinite spins in a quantum anisotropic
Heisenberg $XY$ model. The bath can be treated effectively as a
single pseudo-spin of $N/2$ spin degree. After the
Holstein-Primakoff transformation, it will further be considered as
a single-mode boson at the thermodynamic limit. The pair of qubits
with no direct interaction served as an quantum open subsystem are
initially prepared in a product state or a Bell state. Then the
concurrence of the two qubits, the $z$-component of one of the
subsystem spins and the fidelity of the subsystem can be determined
by a novel polynomial scheme during the temporal evolution. It is
found that (i) larger anisotropic parameter $\gamma$ can help to
maintain the initial state of the two qubits; (ii) the bath at
higher temperature plays a strong destroy effect on the entanglement
and coherence of the subsystem, so does the one with strong
interaction $g_0$; (iii) the dynamics of the subsystem is dependent
on the initial state and in some special cases, only the concurrence
is not sufficient to judge the revival of the subsystem.

\begin{acknowledgments}
We would like to acknowledge the support from the China National
Natural Science Foundation.
\end{acknowledgments}

\end{document}